\documentclass[11pt,epsf]{article}
\usepackage{comment}
\usepackage{amsmath}
\newtheorem{theorem}{Theorem}
\newtheorem{lemma}{Lemma}
\newcommand {\dfn} {\stackrel{\Delta} {=}}

\newcommand {\bu} {\mbox{\boldmath $u$}}

\newcommand {\bx} {\mbox{\boldmath $x$}}
\newcommand {\by} {\mbox{\boldmath $y$}}
\newcommand {\bz} {\mbox{\boldmath $z$}}

\newcommand{\calE}{{\cal E}}

\newcommand{\calI}{{\cal I}}

\newcommand{\calN}{{\cal N}}

\newcommand{\calU}{{\cal U}}

\newcommand{\calX}{{\cal X}}
\newcommand{\calY}{{\cal Y}}
\newcommand{\calZ}{{\cal Z}}
\newcommand{\hP}{\hat{P}}
\newcommand{\hH}{\hat{H}}
\newcommand{\hI}{\hat{I}}
\newcommand{\tP}{\tilde{P}}
\newcommand{\tH}{\tilde{H}}

\allowdisplaybreaks

\begin{document}
\thispagestyle{empty}
\title{Lempel-Ziv Complexity, Empirical Entropies, and Chain Rules}
\author{Neri Merhav}
\date{}
\maketitle

\begin{center}
The Andrew \& Erna Viterbi Faculty of Electrical Engineering\\
Technion - Israel Institute of Technology \\
Technion City, Haifa 32000, ISRAEL \\
E--mail: {\tt merhav@ee.technion.ac.il}\\
\end{center}
\vspace{1.5\baselineskip}
\setlength{\baselineskip}{1.5\baselineskip}

\begin{abstract}
We derive upper and lower bounds on the overall compression ratio of the 1978
Lempel-Ziv (LZ78)
algorithm, applied independently to $k$-blocks of a finite individual
sequence. Both bounds are given in terms of normalized
empirical entropies of the given sequence. For the bounds to be tight and
meaningful, the order of the empirical entropy should be small relative to $k$
in the upper bound, but large relative to $k$ in the lower bound. Several
non-trivial conclusions arise from these bounds. One of them is a certain form
of a chain rule of the Lempel-Ziv (LZ) complexity, which decomposes the joint LZ
complexity of two sequences, say, $\bx$ and $\by$, into the sum of the LZ complexity of $\bx$
and the conditional LZ complexity of $\by$ given $\bx$ (up
to small terms). The
price of this decomposition, however, is in changing the length of the block.
Additional conclusions are discussed as well.
\end{abstract}

\setcounter{section}{0}

\section{Introduction}
\label{intro}

In the second half of the 1970s, Jacob Ziv and Abraham Lempel introduced a transformative concept in information theory 
\cite{ZL77}, \cite{Ziv78}, \cite{ZL78}. Departing from traditional probabilistic frameworks which typically assumed 
memoryless sources and channels with well-defined statistical characteristics, 
they proposed a novel perspective known as the {\em individual-sequence
approach}. This approach, when coupled with models 
of finite-state (FS) encoders and decoders, opened up a new avenue for 
understanding universal data compression and coded communication. Within this innovative framework, 
the foundational ideas of what would become the Lempel-Ziv (LZ) algorithms began to take shape, 
culminating in the development of the LZ77 and LZ78 algorithms in 1977 and
1978, respectively, as well as quite a few other variants.
These algorithms have since become iconic in the field, celebrated not only
for their theoretical elegance, but also for their exceptional 
practical utility. The impact of the LZ family of algorithms,
including its subsequent variations, has been far-reaching, 
deeply embedded in the everyday technologies that rely on digital 
storage and communication, from computers and smartphones to virtually every device that handles digital data.

In subsequent years, the individual-sequence framework was extended along
various lines of study. One notable development 
appeared in \cite{Ziv84}, where Ziv examined a fixed-rate coding scenario involving 
side information, with both the source and the side information sequences
being deterministic (individual) sequences. Along this setting, 
with both the encoder and decoder being modeled as finite-state machines, he introduced and rigorously characterized the 
concept of fixed-rate conditional complexity. 
This measure captures the minimum achievable rate for almost-lossless compression 
of a source sequence given a side information sequence. Remarkably, echoing the classical result from 
Slepian-Wolf coding \cite{SW73}, Ziv demonstrated that access to side information at the encoder is not 
necessary in order to attain this conditional complexity.
The following year, in \cite{Ziv85}, Ziv proposed a variable-rate counterpart to the conditional Lempel-Ziv (LZ) complexity 
in a markedly different context, serving as a universal decoding metric for unknown finite-state channels. This conditional 
complexity measure later garnered attention for its applicability to source coding with side information, 
as explored further in \cite{me00}, \cite{UK03}, and more recently in \cite{meSW}.

Just as LZ complexity serves as the individual-sequence analogue of the entropy rate in the probabilistic setting, 
the conditional LZ complexity naturally parallels the conditional entropy rate. Following this line of analogy between the 
probabilistic and individual-sequence frameworks, a compelling question
arises: Does the LZ complexity measure obey a chain rule? That is, can the joint LZ complexity of a 
sequence pair, say, $(\bx,\by)$, be decomposed into the sum of the LZ
complexity of $\bx$ and
the conditional LZ complexity of $\by$ given $\bx$, or,
symmetrically, the reverse?

On the face of it, a close examination of the mathematical expressions of
these three complexity measures for finite-length 
sequences offers very little reason to hope for an affirmative answer to this question. 
Surprisingly, however, such a chain-rule decomposition 
was shown to hold at least in a specific sense of the asymptotic regime of
infinitely long sequences \cite{meSW}. 
Given the central role that the chain rule for Shannon entropy plays in classical information theory, 
it is natural to envision that an analogous chain rule for LZ complexity could emerge as a foundational 
principle in the development of an information theory tailored to individual sequences.

Consider, for instance, the problem of separately compressing almost losslessly and jointly decompressing two individual 
source sequences, in the spirit of Slepian-Wolf coding \cite{SW73}, but with
the limitation that only finite-state encoders are allowed. As explored in \cite{meSW}, characterizing the 
achievable rate region in this setting brings forth a fundamental question.
In the classical probabilistic framework, for two correlated discrete memoryless sources 
$X$ and $Y$, 
the achievable rate region is well understood. It is defined by the set 
of rate pairs, $\{(R_{\mbox{\tiny x}},R_{\mbox{\tiny
y}}):~R_{\mbox{\tiny x}}\ge H(X|Y),~R_{\mbox{\tiny y}}\ge H(Y|X),~R_{\mbox{\tiny
x}}+R_{\mbox{\tiny y}}\ge H(X,Y)\}$,
where the corner points,
$(H(X),H(Y|X))$ and 
$(H(X|Y),H(Y))$, arise naturally due to the chain rule of entropy, i.e., 
$H(X)=H(X,Y)-H(Y|X)$ and 
$H(Y)=H(X,Y)-H(X|Y)$. 
In the individual-sequence framework, as described in \cite{meSW}, a similar
region can be defined, this time, replacing entropic quantities with their corresponding LZ complexity counterparts: 
the conditional complexities of the sequences and their joint complexity. However, unlike the probabilistic case, 
it is not immediately clear a-priori whether a chain rule exists that allows for a decomposition of the joint LZ complexity 
into unconditional components, such as the individual complexities of 
$\bx$ and $\by$
in a manner analogous to the marginal entropy components of the corner points.
In this context, the existence of a chain rule for LZ complexities would be,
not only natural, but also instrumental in shaping a deeper understanding of compression limits for individual sequences.

Another illustrative example involves the concept 
of successive refinement for individual sequences, as explored in \cite{melast}. 
In the system model considered there, the encoder architecture comprises two
main components: a reproduction encoder and a cascaded
finite-state lossless encoder. The reproduction encoder generates two
distorted versions
of the source, one providing a coarse approximation with relatively high distortion, and the other offering a finer, 
more accurate representation with reduced distortion. These two reproduction vectors are then passed to the lossless encoder, 
which produces two compressed bit-streams that, taken together, 
represent both reproductions without introducing any additional distortion.
The first bit-stream corresponds to the coarse description, 
and ideally, it alone captures the LZ complexity of the coarse reproduction. 
Together, the two bit-streams are expected to match the joint LZ complexity of both reproduction vectors. 
Since the first stream compresses only the coarse version, it is most natural for the second-stage encoder to compress the 
refined reproduction using the coarse one as side information. Consequently, achieving the overall joint LZ complexity hinges 
on the existence of a chain rule for LZ complexity, at least in an asymptotic sense.

Earlier, we mentioned that in \cite{meSW} a certain form of an asymptotic
chain rule of LZ complexities was established for infinite individual
sequences (more details will be provided in Section \ref{chainrule}). Our main result in
this work is another form of a chain rule that applies even to finite
sequences, and it is therefore, stronger, more refined, and more explicit.
To this end, we first derive upper and lower bounds on the overall compression ratio of the 
LZ algorithm, applied independently to $k$-blocks of a finite individual
sequence. Both bounds are given in terms of normalized
empirical entropies of the given sequence. For the bounds to be tight and
meaningful, the order the empirical entropy should be small relative to $k$
in the upper bound, but large relative to $k$ in the lower bound. Several
non-trivial conclusions arise from these bounds. One of them is the above
mentioned chain rule
of the Lempel-Ziv (LZ) complexity, which decomposes the joint LZ
complexity of two sequences into the sum of the LZ complexity of one sequence
and the conditional LZ complexity of the other sequence given the former (up
to small terms). The
price of this decomposition, however, is in changing the length of the block.
Additional conclusions are discussed as well.

Finally, it is interesting to point out that the Kolmogorov complexity
also obeys a certain approximate chain rule (up to a certain redundancy term
that vanishes as the sequence length grows), as asserted in the
Kolmogorov-Levin theorem \cite{Kolmogorov68}, \cite{ZL70}.

The outline of the remaining part of this work is as follows.
In Section \ref{ncbg}, we establish notation conventions and provide some
background. In Section \ref{LZcomplexitybounds}, we derive upper and lower
bounds on the average LZ complexity of $k$-blocks of a given individual sequence, in terms of
empirical entropies, and finally, in Section \ref{chainrule}, we provide the
chain rule results, which are upper and lower bounds on the average of the LZ
complexities of sequence pairs, in terms of the average of the chain-rule
decompositions in a sense that will be made clear in the sequel. We end this
paper with a comparison to the above-mentioned earlier derived chain rule for the LZ complexity
which appears in \cite{meSW}.

\section{Notation Conventions and Background}
\label{ncbg}

\subsection{Notation Conventions}
\label{nc}

Throughout this article, we adopt the following notational conventions. 
Scalar random variables (RVs) will be represented by uppercase letters, their realizations 
by the corresponding lowercase letters, and their alphabets by calligraphic letters. 
The same convention extends to random vectors and their realizations, 
which will be denoted using superscripts to indicate dimension. For instance, 
$X^m$, $m$ being a positive integer,
denotes the random vector $(X_1,\ldots,X_m)$
and $(x_1,\ldots,x_m)$
represents a specific realization in $\calX^m$,
the $m$-fold Cartesian power of the alphabet $\calX$.
Segment notation will follow accordingly: for positive integers $i$ and $j$,
$i\le j$, $x_i^j$ and $X_i^j$ denote
the substrings $(x_i,x_{i+1},\ldots,x_j)$ and
$(X_i,X_{i+1},\ldots,X_j)$,
respectively. When $i=1$
the subscript `1' is omitted for brevity. If
$i>j$, both 
$x_i^j$ and $X_i^j$
refer to the empty string.
Unless stated otherwise, all logarithms and exponentials are taken to base 2.
The indicator function of an event $\calE$ is denoted by $I\{\calE\}$, that
is, $I\{\calE\}=1$ if $\calE$ occurs and
$I\{\calE\}=0$ if not.

In the sequel, $x^n=(x_1,\ldots,x_n)$ and
$y^n=(y_1,\ldots,y_n)$ and
will designate individual sequences.
The components, $\{x_i\}$ of $x^n$, and $\{y_i\}$ of $y^n$, all take values in
the corresponding finite
alphabets, $\calX$ and $\calY$, whose cardinalities will be denoted by
$\alpha$ and $\beta$, respectively.
The infinite sequences $(x_1,x_2,\ldots)$ and
$(y_1,y_2,\ldots)$ and
will be denoted by $\bx$ and $\by$, respectively.

\subsection{Background}
\label{bg}

\subsubsection{Finite-State Encoders}

Following the framework introduced in \cite{ZL78}, we consider a model for lossless 
compression based on a finite-state encoder. Such an encoder is characterized by a quintuple
$$E=(\calX,\calU,\calZ,f,g),$$
where:
$\calX$ denotes a finite input alphabet of cardinality $\alpha = |\calX|$;
$\calU$ is a finite set of variable-length binary strings, possibly including the empty string $\lambda$ (of length zero);
$\calZ$ is a finite set of internal states;
$f: \calZ \times \calX \to \calU$ is the output function, and
$g: \calZ \times \calX \to \calZ$ is the next-state transition function.
Given an infinite input sequence $\bx = (x_1, x_2, \ldots)$ with $x_i \in
\calX$, $i=1,2,\ldots$, henceforth referred to as the source sequence, the encoder $E$ generates a 
corresponding infinite output sequence $\bu = (u_1, u_2, \ldots)$ with $u_i \in \calU$,
henceforth termed the compressed bit-stream, while simultaneously evolving through a sequence of internal 
states $\bz = (z_1, z_2, \ldots)$ with $z_i \in \calZ$. The system dynamics are governed recursively by the equations:
\begin{eqnarray}
u_i &=& f(z_i, x_i), \label{yi} \\
z_{i+1}&=&g(z_i,x_i), \label{nextstate}
\end{eqnarray}
for $i = 1, 2, \ldots$, with a fixed initial state $z_1 = z_\star\in\calZ$. 
If at any step $u_i = \lambda$, no output is produced, and this corresponds to encoder idling, 
where only the internal state is updated in response to the input symbol.

An encoder with $s$ distinct internal states, henceforth referred to as an $s$-state
encoder, is one for which $|\calZ| = s$.
For convenience, we adopt a few notational conventions from \cite{ZL78}:
Given a segment of input symbols $x_i^j$ with $i \le j$ and an initial state $z_i$, 
the notation $f(z_i, x_i^j)$ denotes the corresponding segment of outputs $u_i^j$ generated by the encoder $E$. 
Likewise, $g(z_i, x_i^j)$ denotes the resulting state $z_{j+1}$ after processing the 
input segment $x_i^j$ starting from state $z_i$.

A finite-state encoder $E$ is said to be information lossless (IL) if, 
for every initial state $z_i \in \calZ$, every positive integer $n$, and any
input segment $x_i^{i+n}$, the triplet 
$(z_i,f(z_i,x_i^{i+n}),g(z_i,x_i^{i+n}))$
uniquely determines the original input segment $x_i^{i+n}$. In other words, 
the combination of the starting state, the resulting output sequence, 
and the final state after encoding is sufficient to fully reconstruct the input.
Given an encoder $E$ and an input sequence $x^n$, the compression ratio achieved by $E$ on $x^n$ is defined as
\begin{equation}
\rho_E(x^n)\dfn \frac{L(u^n)}{n}=\frac{1}{n}\sum_{i=1}^nl(u_i)=\frac{1}{n}\sum_{i=1}^nl[f(z_i,x_i)]
\end{equation}
where $L(u^n)$ denotes the total length (in bits) of the encoded binary string $u^n$, 
and $l(u_i)$ represents the length of the binary string $u_i = f(z_i, x_i)$ at each step $i$.

The class of all IL encoders $\{E\}$ with no more than $s$
states is denoted by $\calE(s)$. We next define the $s$-state compressibility
of $x^n$ by
\begin{equation}
\rho_s(x^n)=\min_{E\in \calE(s)}\rho_E(x^n),
\end{equation}
the asymptotic $s$-state compressibility of $\bx$ by
\begin{equation}
\rho_s(\bx)=\limsup_{n\to\infty}\rho_s(x^n),
\end{equation}
and finally, the {\it finite--state compressibility} of $\bx$ by
\begin{equation}
\rho(\bx)=\lim_{s\to\infty}\rho_s(\bx).
\end{equation}

\subsubsection{Empirical Distributions and Induced Information Measures}

We define three types of empirical distributions of $d$-vectors ($d$ --
positive integer). 
\begin{enumerate}
\item Assuming that $d$ divides $n$, the empirical distribution pertaining to {\em
non-overlapping blocks} of length $d$ is defined as
\begin{equation}
\hP_{\mbox{\tiny
nob}}(z,w^d)\dfn\frac{d}{n}\sum_{i=0}^{n/d-1}\calI\{z_{id+1}=z,~x_{id+1}^{id+d}=w^d\},~~~
z\in\calZ,~w^d=(w_1,\ldots,w_d)\in\calX^d.
\end{equation}
\item The empirical distribution associated with a {\em sliding window} of
length $d$ is defined as
\begin{equation}
\hP_{\mbox{\tiny
sw}}(z,w^d)\dfn\frac{1}{n-d+1}\sum_{i=0}^{n-d}\calI\{z_{i+1}=z,~x_{i+1}^{i+d}=w^d\},~~~z\in\calZ,~w^d\in\calX^d.
\end{equation}
\item The empirical distribution associated with a {\em cyclic sliding window} of
length $d$ is defined as
\begin{equation}
\hP_{\mbox{\tiny
csw}}(z,w^d)\dfn\frac{1}{n}\sum_{i=0}^{n-1}\calI\{z_{i+1}=z,~x_{i+1}^{[(i-1)\oplus
d]+1}=w^d\},~~~z\in\calZ,~w^d\in\calX^d,
\end{equation}
where $\oplus$ denotes modulo-$n$ addition.
\end{enumerate}

Information measures associated with these empirical distributions will be
denoted according to the conventional notation rules of the information theory
literature, but with `hats', with subscripts that indicate the type of the
empirical distribution, and with notation of dependence on the data sequence
$x^n$ from which the statistics were gathered (using square brackets).
For example, $\hH_{\mbox{\tiny nob}}(X^d)[x^n]$ will denote the
empirical entropy of an auxiliary random vector $X^d$ that is governed by
the empirical distribution, $\hP_{\mbox{\tiny nob}}(\cdot)$ extracted from
$x^n$.\footnote{Note that there is no need to denote the dependence on $z^n$
too since $z^n$ is dictated by $x^n$ for a given next-state function, $g$.} Likewise,
$\hH_{\mbox{\tiny sw}}(X^d|Z)[x^n]$ will denote the
empirical conditional entropy of an auxiliary random vector $X^d$ given a
random state variable $Z$ that are drawn by
the empirical distribution, $\hP_{\mbox{\tiny sw}}(\cdot,\cdot)$,
$\hI_{\mbox{\tiny csw}}(X^d;Z)[x^n]$ will denote the
empirical mutual information between the auxiliary random variables $X^d$ and
$Z$ that are jointly disributed according to
the empirical distribution, $\hP_{\mbox{\tiny csw}}(\cdot)$, and so on.

For the infinite sequence $\bx=(x_1,x_2,\ldots)$, we define
\begin{equation}
\hH_{\mbox{\tiny csw}}(X^d)[\bx]=\limsup_{n\to\infty}
\hH_{\mbox{\tiny csw}}(X^d)[x^n].
\end{equation}
Similarly as shown in \cite{ZL78}, for every $\bx$, the sequence $\{\hH_{\mbox{\tiny
csw}}(X^d)[\bx]\}_{d\ge 1}$ is sub-additive as
\begin{eqnarray}
\hH_{\mbox{\tiny csw}}(X^{d_1+d_2})[x^n]&=&
\hH_{\mbox{\tiny csw}}(X^{d_1})[x^n]+
\hH_{\mbox{\tiny csw}}(X_{d_1+1}^{d_1+d_2}|X^{d_1})[x^n]\nonumber\\
&\le&\hH_{\mbox{\tiny csw}}(X^{d_1})[x^n]+
\hH_{\mbox{\tiny csw}}(X_{d_1+1}^{d_1+d_2})[x^n]\nonumber\\
&=&\hH_{\mbox{\tiny csw}}(X^{d_1})[x^n]+
\hH_{\mbox{\tiny csw}}(X^{d_2})[x^n],
\end{eqnarray}
and so, taking the limit superior of the left-most- and the right-most side,
we have 
\begin{eqnarray}
\hH_{\mbox{\tiny csw}}(X^{d_1+d_2})[\bx]&=&
\limsup_{n\to\infty}\hH_{\mbox{\tiny csw}}(X^{d_1+d_2})[x^n]\nonumber\\
&\le&\limsup_{n\to\infty}\{\hH_{\mbox{\tiny csw}}(X^{d_1})[x^n]+
\hH_{\mbox{\tiny csw}}(X^{d_2})[x^n]\}\nonumber\\
&\le&\limsup_{n\to\infty}\hH_{\mbox{\tiny csw}}(X^{d_1})[x^n]+
\limsup_{n\to\infty}\hH_{\mbox{\tiny csw}}(X^{d_2})[x^n]\nonumber\\
&=&\hH_{\mbox{\tiny csw}}(X^{d_1})[\bx]+
\hH_{\mbox{\tiny csw}}(X^{d_2})[\bx].
\end{eqnarray}
Consequently, the sequence
$\{\frac{\hH_{\mbox{\tiny csw}}(X^d)[\bx]}{d}\}_{d\ge 1}$ is convergent, and
we shall denote 
\begin{equation}
\bar{H}_{\mbox{\tiny csw}}[\bx]=\lim_{d\to\infty} \frac{\hH_{\mbox{\tiny
csw}}(X^d)[\bx]}{d}.
\end{equation}

Returning to finite $n$,
whenever the underlying sequence $x^n$ is clear from the context, we will omit
the explicit notation that indicates the dependence upon $x^n$. In this case,
the above-mentioned examples of information measures will be denoted more simply by 
$\hH_{\mbox{\tiny nob}}(X^d)$, $\hH_{\mbox{\tiny sw}}(X^d|Z)$, and
$\hI_{\mbox{\tiny csw}}(X^d;Z)$, respectively.

\subsubsection{LZ Compression and its Properties}

The incremental parsing procedure used in the LZ78 algorithm is a sequential method 
for processing an input sequence $x^n$ drawn from a finite alphabet. 
At each step, the procedure identifies the shortest substring that has not yet appeared as a complete phrase 
in the current parsed set except possibly for the final (incomplete) phrase.
For instance, applying this parsing method to the sequence
$$x^{15}=\mbox{abbabaabbaaabaa}$$ 
yields
$$\mbox{a,b,ba,baa,bb,aa,ab,aa}.$$ 
Let $c(x^n)$ denote the total number of distinct phrases generated by this procedure 
(in this example, $c(x^{15}) = 8$). Additionally, let $LZ(x^n)$ represent the length 
in bits of the binary string produced by the LZ78 encoding of $x^n$.
According to Theorem 2 of \cite{ZL78}, the following inequality holds:
\begin{equation}
\label{lz-clogc}
LZ(x^n)\le[c(x^n)+1]\log\{2\alpha[c(x^n)+1]\}
\end{equation}
which can easily be shown to be further upper bounded by
\begin{equation}
\label{epsilon1}
LZ(x^n)\le c(x^n)\log c(x^n)+n\cdot\epsilon_1(n),
\end{equation}
where $\epsilon_1(n)$ tends to zero uniformly as $n\to\infty$.
In other words, the LZ78 code length for a sequence $x^n$ is upper bounded by an expression whose 
leading term is $c(x^n)\log c(x^n)$. Remarkably, the very same quantity also appears as the dominant term in a 
lower bound (see Theorem 1 of \cite{ZL78}) on the shortest code length achievable by any 
IL finite-state encoder with no more than $s$ states, assuming that $\log(s^2)$ is negligible in 
comparison to $\log c(x^n)$.
More precisely, Theorem 1 in \cite{ZL78} asserts that:
\begin{equation}
\rho_s(x^n)\ge\frac{c(x^n)+s^2}{n}\cdot\log\left(\frac{c(x^n)+s^2}{4s^2}\right)+\frac{2s^2}{n},
\end{equation}
which can readily be further lower bounded by
\begin{equation}
\label{epsilon2}
\rho_s(x^n)\ge
\frac{c(x^n)\log c(x^n)}{n}-\epsilon_2(n,s),
\end{equation}
where $\epsilon_2(n,s)\to 0$ uniformly as $n\to\infty$ for fixed $s$.
Motivated by this connection, we shall refer to the quantity 
$c(x^n)\log c(x^n)$ as
the unnormalized LZ complexity of $x^n$ 
The normalized LZ complexity is then defined as
\begin{equation}
\rho_{\mbox{\tiny LZ}}(x^n)\dfn
\frac{c(x^n)\log
c(x^n)}{n},
\end{equation}
which represents the LZ complexity per input symbol.

\section{Bounds on the Average LZ Complexity of $k$-Blocks}
\label{LZcomplexitybounds}

In this section, we derive lower and upper bounds on the average LZ complexity
over blocks of length $k$, that is, on the quantity
\begin{equation}
\frac{k}{n}\sum_{i=0}^{n/k-1}\rho_{\mbox{\tiny LZ}}(x_{ik+1}^{ik+k})=
\frac{1}{n}\sum_{i=0}^{n/k-1}c(x_{ik+1}^{ik+k})\log c(x_{ik+1}^{ik+k}),
\end{equation}
where $k$ is a positive integer that divides $n$. Both the upper bound and the
lower bound are given in terms of the empirical entropy $\hH_{\mbox{\tiny
csw}}(\cdot)$, but to make certain redundancy terms negligibly small, the order of this empirical entropy
should be much larger than $k$ in the lower bound and
much smaller than $k$ in the upper bound.

The reason for our interest
in the average LZ complexity of blocks, rather than in the LZ complexity of
the entire sequence, $\rho_{\mbox{\tiny LZ}}(x^n)$, is that in any
practical application of the LZ78 algorithm, one must reset and start over
after each and every block of finite size, as otherwise, the amount of memory and
computational effort grows without bound. Also, from the theoretical point of
view (see \cite[Corollary 2]{ZL78}), the gap between the upper bound and the lower bound to the finite-state
complexity of $\bx$ is closed in the limit of $s\to\infty$, in terms of the double limit 
\begin{equation}
\rho(\bx)=\limsup_{k\to\infty}
\limsup_{n\to\infty}\frac{k}{n}\sum_{i=0}^{n/k-1}\rho_{\mbox{\tiny
LZ}}(x_{ik+1}^{ik+k}).
\end{equation}
One of the conclusions from our bounds in this section is that, in fact, the
outer limit superior over $k$ can be always safely replaced by an ordinary limit,
because the sequence
\begin{equation}
\rho_k(\bx)\dfn
\limsup_{n\to\infty}\frac{k}{n}\sum_{i=0}^{n/k-1}\rho_{\mbox{\tiny
LZ}}(x_{ik+1}^{ik+k}),~~~~k\in\calN,
\end{equation}
turns out to be convergent thanks to the convergence of the sequence $\{\frac{\hH_{\mbox{\tiny
csw}}(X^d)[\bx]}{d}\}_{d\ge 1}$, which plays a role both in the upper bound
and in the lower bound, as described above.

\subsection{Lower Bound}
\label{lowerbound}

The following theorem provides our lower bound to the average of the LZ complexities
over $k$-blocks in terms of the cyclic sliding-window empirical entropy.

\begin{theorem}
\label{lb}
For every positive integer $k$, every $n$ that is an integer multiple of $k$,
every $x^n\in\calX^n$, and every positive integer $\ell$,
\begin{equation}
\label{Hlowerbound}
\frac{k}{n}\sum_{i=0}^{n/k-1}\rho_{\mbox{\tiny
LZ}}(x_{ik+1}^{ik+k})\ge \frac{\hH_{\mbox{\tiny csw}}(X^\ell)}{\ell}-
\Delta_\ell(\alpha^{k},\alpha)-\frac{\ell\log\alpha}{n}-\frac{\alpha^\ell}{n}\log \frac{n}{\ell}-\epsilon_1(k),
\end{equation}
where $\epsilon_1(\cdot)$ is as in (\ref{epsilon1}) and
\begin{equation}
\Delta_\ell(s,\alpha)\dfn\frac{1}{\ell}\log\left\{s^2\cdot\left[1+
\log\left(1+\frac{\alpha^\ell}{s}\right)\right]\right\}.
\end{equation}
\end{theorem}

Since the left-hand side of (\ref{Hlowerbound}) does not depend on $\ell$, in principle, one 
could maximize the right-hand side over $\ell$ to obtain the tightest lower
bound. But perhaps a more  natural point of view is to consider the regime
$n\gg\ell\gg k\gg 1$, where the leading term of the lower bound is the
empirical entropy term and all other four terms are negligibly small. In
particular, taking the limit superior $n\to\infty$, followed by the limit
$\ell\to\infty$, and finally, the limit inferior $k\to\infty$, we
obtain the following asymptotic inequality as a consequence of eq.\
(\ref{Hlowerbound}):
\begin{equation}
\label{rhogeH}
\liminf_{k\to\infty}\rho_k(\bx)\ge \bar{H}_{\mbox{\tiny csw}}[\bx].
\end{equation}
For later use, we point out that the lower bound of Theorem \ref{lb} can also
be expressed in terms of the empirical entropy associated with non-overlapping
blocks, in the following manner:
\begin{equation}
\label{Hlowerboundnob}
\frac{k}{n}\sum_{i=0}^{n/k-1}\rho_{\mbox{\tiny
LZ}}(x_{ik+1}^{ik+k})\ge \frac{\hH_{\mbox{\tiny nob}}(X^\ell)}{\ell}-
\Delta_\ell(\alpha^{k},\alpha)-\epsilon_1(k),
\end{equation}
whose proof is very similar to (and even slightly simpler than) 
the proof of Theorem \ref{lb} below. This will be used in Section
\ref{chainrule}.

The remaining part of this subsection is devoted to the proof of Theorem
\ref{lb}.\\

\noindent
{\em Proof of Theorem \ref{lb}.}
We commence by providing a generalized version of
Kraft's inequality that applies to any $s$-state IL encoder.
It is similar but somewhat different (and slightly tighter) than the generalized Kraft
inequality of
\cite[Lemma 2]{ZL78}. 

\begin{lemma}
\label{gkraft}
For every IL encoder with $s$ states and every $z\in\calZ$,
\begin{equation}
K(z)\dfn\sum_{w^\ell\in\calX^\ell}2^{-L[f(z,w^\ell)]}\le
s\cdot\left[1+\log\left(1+\frac{\alpha^\ell}{s}\right)\right].
\end{equation}
\end{lemma}

The proof of Lemma \ref{gkraft} is identical to the proof of Lemma 2 of \cite{ZL78} except that since
the initial state $z$ is given and fixed, the number $k_j$ of different $\{w^\ell\}$ with
$L[f(z,w^\ell)]=j$ cannot exceed $s\cdot 2^j$ (rather than $s^2 2^j$ in
\cite{ZL78}), which is the number of combinations of final states and binary
output sequences of length $j$.

Next, observe that
\begin{eqnarray}
& &s\cdot\left[1+\log\left(1+\frac{\alpha^\ell}{s}\right)\right]\nonumber\\
&\ge&\sum_{w^\ell\in\calX^\ell}2^{-L[f(z,w^\ell)]}\nonumber\\
&=&\sum_{w^\ell\in\calX^\ell}\hP_{\mbox{\tiny sw}}(w^\ell|z)\cdot
2^{-L[f(z,w^\ell)]-\log\hP_{\mbox{\tiny sw}}(w^\ell|z)}\nonumber\\
&\ge&\exp_2\left\{-\sum_{w^\ell\in\calX^\ell}\hP_{\mbox{\tiny sw}}(w^\ell|z)\cdot
L[f(z,w^\ell)]-\sum_{w^\ell}\hP_{\mbox{\tiny sw}}(w^\ell|z)\log\hP_{\mbox{\tiny sw}}(w^\ell|z)\right\}\nonumber\\
&=&\exp_2\left\{\hH_{\mbox{\tiny sw}}(X^\ell|Z=z)-\sum_{w^\ell\in\calX^\ell}\hP_{\mbox{\tiny sw}}(w^\ell|z)\cdot
L[f(z,w^\ell)]\right\},
\end{eqnarray}
where in the second inequality we have used Jensen's inequality and the
convexity of the exponential function, $F(u)=2^u$. This
implies that
\begin{equation}
\sum_{w^\ell\in\calX^\ell}\hP_{\mbox{\tiny sw}}(w^\ell|z)\cdot L[f(z,w^\ell)]\ge
\hH_{\mbox{\tiny sw}}(X^\ell|Z=z)-\log\left\{s\cdot\left[1+\log\left(1+\frac{\alpha^\ell}{s}\right)\right]\right\}.
\end{equation}
Averaging both sides w.r.t. $\hat{P}_{\mbox{\tiny sw}}(z)$, $z\in\calZ$, we end up with
\begin{equation}
\label{empconv}
\sum_{(z,w^\ell)\in\calZ\times\calX^\ell}\hP_{\mbox{\tiny sw}}(z,w^\ell)\cdot L[f(z,w^\ell)]\ge
\hH_{\mbox{\tiny sw}}(X^\ell|Z)-\log\left\{s\cdot\left[1+\log\left(1+\frac{\alpha^\ell}{s}\right)\right]\right\}.
\end{equation}
We next apply this inequality in a chain of inequalities that would lead to a
lower bound to $\rho_E(x^n)$. 
Similarly as in eq.\ (33) of \cite{ZL78}
\begin{eqnarray}
\rho_E(x^n)
&=&\frac{1}{n}\sum_{i=1}^nl[f(z_i,x_i)]\nonumber\\
&=&\frac{1}{n\ell}\sum_{i=1}^n\ell\cdot l[f(z_i,x_i)]\nonumber\\
&\ge&\frac{1}{n\ell}\sum_{i=0}^{n-\ell}\sum_{j=1}^\ell l[f(z_{i+j},
x_{i+j})]\nonumber\\
&=&\frac{1}{n\ell}\sum_{i=0}^{n-\ell}
L[f(z_{i+1},x_{i+1}^{i+\ell})]\nonumber\\
&=&\frac{1}{\ell}\left(1-\frac{\ell-1}{n}\right)\cdot\sum_{z,w^\ell}\hP_{\mbox{\tiny sw}}(z,w^\ell)\cdot
L[f(z,w^\ell)]\nonumber\\
&\ge&\left(1-\frac{\ell}{n}\right)\cdot\frac{\hH_{\mbox{\tiny sw}}(X^\ell|Z)}{\ell}-
\frac{1}{\ell}\log\left\{s\cdot\left[1+\log\left(1+\frac{\alpha^\ell}{s}\right)\right]\right\}\nonumber\\
&=&\left(1-\frac{\ell}{n}\right)\cdot\frac{\hH_{\mbox{\tiny sw}}(X^\ell)-\hat{I}_{\mbox{\tiny sw}}(Z;X^\ell)}{\ell}-
\frac{1}{\ell}\log\left\{s\cdot\left[1+\log\left(1+\frac{\alpha^\ell}{s}\right)\right]\right\}\nonumber\\
&\ge&\left(1-\frac{\ell}{n}\right)\cdot\frac{\hH_{\mbox{\tiny sw}}(X^\ell)-\hat{H}_{\mbox{\tiny sw}}(Z)}{\ell}-
\frac{1}{\ell}\log\left\{s\cdot\left[1+\log\left(1+\frac{\alpha^\ell}{s}\right)\right]\right\}\nonumber\\
&\ge&\left(1-\frac{\ell}{n}\right)\cdot\frac{\hH_{\mbox{\tiny sw}}(X^\ell)-\log s}{\ell}-
\frac{1}{\ell}\log\left\{s\cdot\left[1+\log\left(1+\frac{\alpha^\ell}{s}\right)\right]\right\}\nonumber\\
&\ge&\frac{\hH_{\mbox{\tiny sw}}(X^\ell)}{\ell}-\frac{\ell\log\alpha}{n}-
\frac{1}{\ell}\log\left\{s^2\cdot\left[1+\log\left(1+\frac{\alpha^\ell}{s}\right)\right]\right\}.
\end{eqnarray}
We would now like to modify this lower bound to be given in terms of the
empirical entropy
$\hH_{\mbox{\tiny csw}}(X^\ell)$. 
It is easy to verify that
$|\hP_{\mbox{\tiny csw}}(w^\ell)-\hP_{\mbox{\tiny
sw}}(w^\ell)|\le\frac{\ell}{n}$ for all $w^\ell\in\calX^\ell$, and so, the
variational distance between $\hP_{\mbox{\tiny csw}}(\cdot)$ and
$\hP_{\mbox{\tiny sw}}(\cdot)$ cannot exceed $\theta\dfn \ell\alpha^\ell/n$.
Thus, by \cite[Lemma 2.7, p.\ 19]{CK11},
\begin{equation}
\hH_{\mbox{\tiny sw}}(X^\ell)\ge
\hH_{\mbox{\tiny
csw}}(X^\ell)-\frac{\ell\alpha^\ell}{n}\log\frac{n}{\ell},
\end{equation}
and so, we have proved that
\begin{equation}
\rho_s(x^n)\ge\frac{\hH_{\mbox{\tiny
csw}}(X^\ell)}{\ell}-\Delta_\ell(s,\alpha)-\frac{\ell\log\alpha}{n}-\frac{\alpha^\ell}{n}\log\frac{n}{\ell}.
\end{equation}
Consider now the application of the LZ78 algorithm along blocks of length
$k$, where after each such block the LZ algorithm is restarted independently
of previous blocks. Since this is actually a block code of block length $k$
and a block code can be implemented using a finite-state encoder with
$s=\alpha^k$ states, then we have:
\begin{eqnarray}
\frac{1}{n}\sum_{i=0}^{n/k-1}LZ(x_{ik+1}^{ik+k})
&\ge&\rho_{\alpha^k}(x^n)\nonumber\\
&\ge&\frac{\hH_{\mbox{\tiny
csw}}(X^\ell)}{\ell}-\Delta_\ell(\alpha^k,\alpha)-\frac{\ell\log\alpha}{n}-\frac{\alpha^\ell}{n}\log
\frac{n}{\ell}.
\end{eqnarray}
On the other hand,
\begin{eqnarray}
\frac{1}{n}\sum_{i=0}^{n/k-1}LZ(x_{ik+1}^{ik+k})
&=&\frac{k}{n}\sum_{i=0}^{n/k-1}\frac{LZ(x_{ik+1}^{ik+k})}{k}\nonumber\\
&\le&\frac{k}{n}\sum_{i=0}^{n/k-1}\rho_{\mbox{\tiny
LZ}}(x_{ik+1}^{ik+k})+\epsilon_1(k),
\end{eqnarray}
and so,
\begin{equation}
\frac{k}{n}\sum_{i=0}^{n/k-1}\rho_{\mbox{\tiny
LZ}}(x_{ik+1}^{ik+k})\ge \frac{\hH_{\mbox{\tiny csw}}(X^\ell)}{\ell}-
\Delta_\ell(\alpha^{k},\alpha)-\frac{\ell\log\alpha}{n}-\frac{\alpha^\ell}{n}\log \frac{n}{\ell}-\epsilon_1(k),
\end{equation}
thus completing the proof of Theorem \ref{lb}.

\subsection{Upper Bound}
\label{upperbound}

Theorem \ref{ub} below provides an upper bound to
$$\frac{k}{n}\sum_{i=0}^{n/k-1}\rho_{\mbox{\tiny
LZ}}(x_{ik+1}^{ik+k}).$$

\begin{theorem}
\label{ub}
For every positive integer $k$, every $n$ that is an integer multiple of $k$,
every $x^n\in\calX^n$, and every positive integer $m$,
\begin{equation}
\label{Hupperbound}
\frac{k}{n}\sum_{i=0}^{n/k-1}\rho_{\mbox{\tiny
LZ}}(x_{ik+1}^{ik+k})\le 
\frac{\hH_{\mbox{\tiny
csw}}(X^{m})}{m}+\frac{1}{m}+\frac{2(m+1)\alpha^{m+1}}{n}\log\frac{n}{m}+\epsilon_2(k,\alpha^{2m}),
\end{equation}
where $\epsilon_2(\cdot,\cdot)$ is as in eq.\ (\ref{epsilon2}).
\end{theorem}

Similarly as in the discussion after Theorem \ref{lb},
since the left-hand side does not depend on $m$, in principle, one
could minimize the right-hand side over $m$ to obtain the tightest upper
bound, but it may be more instructive to consider the regime
$n\gg m\gg 1$ and $k\gg m$, where the leading term of the upper bound is the
empirical entropy term and all other three terms are negligibly small. In
particular, taking the limit superior $n\to\infty$, followed by the limit
superior of $k\to\infty$, and finally, the limit $m\to\infty$, we
obtain the following asymptotic inequality as a consequence of eq.\
(\ref{Hupperbound}):
\begin{equation}
\label{rholeH}
\limsup_{k\to\infty}\rho_k(\bx)\le \bar{H}_{\mbox{\tiny csw}}[\bx],
\end{equation}
which together with eq.\ (\ref{rhogeH}), yields
\begin{equation}
\limsup_{k\to\infty}\rho_k(\bx)=\liminf_{k\to\infty}\rho_k(\bx)=\lim_{k\to\infty}\rho_k(\bx)=\bar{H}_{\mbox{\tiny
csw}}[\bx],
\end{equation}
in agreement with Theorem 3 of \cite{ZL78}, but with the stronger statement
that the limit superior over $k$ is actually an ordinary limit, as the sequence
$\{\rho_k(\bx)\}_{k\ge 1}$ is convergent.

Similarly as before, the upper bound of Theorem \ref{ub} can also
be expressed in terms of the empirical entropy associated with non-overlapping
blocks, in the following manner:
\begin{equation}
\label{Hupperboundnob}
\frac{k}{n}\sum_{i=0}^{n/k-1}\rho_{\mbox{\tiny
LZ}}(x_{ik+1}^{ik+k})\le 
\frac{\hH_{\mbox{\tiny
nob}}(X^{m})}{m}+\frac{1}{m}+\epsilon_2(k,\alpha^{2m}),
\end{equation}
and once again, the proof is almost identical to
the proof of Theorem \ref{ub} below. This result too will be used in Section
\ref{chainrule}.

The remaining part of this subsection is devoted to the proof of Theorem
\ref{ub}.\\

\noindent
{\em Proof of Theorem \ref{ub}.}
Consider a scenario of compressing $x^n$ by a finite-state encoder with
$s$ states that
is allowed to vary from one $k$-block to another. According to eq.\
(\ref{epsilon2}) (applied to $k$-blocks), the corresponding compression ratio,
which is $\frac{k}{n}\sum_{i=0}^{n/k-1}\rho_s(x_{ik+1}^{ik+k})$, is lower
bounded by
\begin{equation}
\frac{k}{n}\sum_{i=0}^{n/k-1}\rho_s(x_{ik+1}^{ik+k})\ge
\frac{k}{n}\sum_{i=0}^{n/k-1}\rho_{\mbox{\tiny
LZ}}(x_{ik+1}^{ik+k})-\epsilon_2(k,s).
\end{equation}
On the other hand, the best time-varying finite-state encoder with
$s=\alpha^{2m}$ states ($m$- positive integer)
cannot be worse than the best time-invariant finite-state encoder with the same number of
states. Let $m$ divide $n$ and consider block encoding of non-overlapping
blocks of length $m$ using a Shannon code with a conditional length function
defined by
\begin{equation}
L(x_{im+1}^{im+m}|x_{(i-1)m+1}^{im})=\Bigg\lceil -\log\bigg[\prod_{j=1}^m
Q(x_{im+j}|x_{(i-1)m+j}^{im+j-1})\bigg]\Bigg\rceil,~~~~i=0,1,2,\ldots,\frac{n}{m}-1
\end{equation}
where for $i=0$, $x_1^m$ is understood to be encoded with fixed arbitrary conditioning on
$x_{-(m-1)}^0\in\calX^m$.
Next, observe that
\begin{eqnarray}
L(x^n|x_{-(m-1)}^0)&\dfn&\sum_{i=0}^{n/m-1}L(x_{im+1}^{im+m}|x_{(i-1)m+1}^{im})\nonumber\\
&=&\sum_{i=0}^{n/m-1}\Bigg\lceil -\log\bigg[\prod_{j=1}^m
Q(x_{im+j}|x_{(i-1)m+j}^{im+j-1})\bigg]\Bigg\rceil\nonumber\\
&\le&-\sum_{i=0}^{n/m-1}\log\left[\prod_{j=1}^m
Q(x_{im+j}|x_{(i-1)m+j}^{im+j-1})\right]+\frac{n}{m}\nonumber\\
&=&-\sum_{i=0}^{n/m-1}\sum_{j=1}^m\log
Q(x_{im+j}|x_{(i-1)m+j}^{im+j-1})+\frac{n}{m}\nonumber\\
&=&-\sum_{i=1}^n\log
Q(x_i|x_{i-m}^{i-1})+\frac{n}{m}\nonumber\\
&=&-n\cdot\sum_{w^{m+1}\in\calX^{m+1}}\tP_{\mbox{\tiny sw}}(w^{m+1})\log Q(w_{m+1}|w^m)+\frac{n}{m},
\end{eqnarray}
where
\begin{equation}
\tP_{\mbox{\tiny sw}}(w^{m+1})=\frac{1}{n}\sum_{i=1}^n\calI\{x_{i-m}^i =
w^{m+1}\},~~~~w^{m+1}\in\calX^{m+1},
\end{equation}
and in the sequel, we denote the induced entropy by $\tH_{\mbox{\tiny
sw}}(\cdot)$. Thus,
\begin{equation}
\frac{k}{n}\sum_{i=0}^{n/k-1}\rho_{\alpha^{2m}}(x_{ik+1}^{ik+k})\le
-\sum_{w^{m+1}\in\calX^{m+1}}\tP_{\mbox{\tiny sw}}(w^{m+1})\log Q(w_{m+1}|w^m)+\frac{1}{m},
\end{equation}
and since this holds for every $Q(\cdot|\cdot)$, it also holds for the
minimizing $Q(\cdot|\cdot)$, which yields
\begin{equation}
\frac{k}{n}\sum_{i=0}^{n/k-1}\rho_{\alpha^{2m}}(x_{ik+1}^{ik+k})\le
\tH_{\mbox{\tiny sw}}(X_{m+1}|X^m)
+\frac{1}{m}.
\end{equation}
At this point, we wish to pass from $\tH_{\mbox{\tiny sw}}(X_{m+1}|X^m)$ to
$\hH_{\mbox{\tiny csw}}(X_{m+1}|X^m)$, as before. To this end, we first present
$\tH_{\mbox{\tiny sw}}(X_{m+1}|X^m)$ as $\tH_{\mbox{\tiny
sw}}(X^{m+1})-\tH_{\mbox{\tiny sw}}(X^m)$ and then apply again Lemma 2.7 of
\cite{CK11} to
each term separately. Since $|\tP_{\mbox{\tiny sw}}(w^m)-\hP_{\mbox{\tiny
csw}}(w^m)|\le \frac{m}{n}$ for all $w^m\in\calX^m$ then,
$\sum_{w^m}|\tP_{\mbox{\tiny sw}}(w^m)-\hP_{\mbox{\tiny csw}}(w^m)|\le
\frac{m\alpha^m}{n}$, and so, by Lemma 2.7 of \cite{CK11},
\begin{equation}
|\tH_{\mbox{\tiny sw}}(X^m)-\hH_{\mbox{\tiny csw}}(X^m)|\le
\frac{m\alpha^m}{n}\log\frac{n}{m}
\end{equation}
and a similar inequality holds also for $|\tH_{\mbox{\tiny
sw}}(X^{m+1})-\hH_{\mbox{\tiny csw}}(X^{m+1})|$.
It follows that
\begin{eqnarray}
\tH_{\mbox{\tiny sw}}(X_{m+1}|X^m)&=&\tH_{\mbox{\tiny
sw}}(X^{m+1})-\tH_{\mbox{\tiny sw}}(X^m)\nonumber\\
&\le&\hH_{\mbox{\tiny
csw}}(X^{m+1})+\frac{(m+1)\alpha^{m+1}}{n}\log\frac{n}{m+1}
-\left[\hH_{\mbox{\tiny
csw}}(X^m)-\frac{m\alpha^m}{n}\log\frac{n}{m}\right]\nonumber\\
&\le&\hH_{\mbox{\tiny
csw}}(X_{m+1}|X^m)+\frac{2(m+1)\alpha^{m+1}}{n}\log\frac{n}{m},
\end{eqnarray}
and so,
\begin{eqnarray}
\frac{k}{n}\sum_{i=0}^{n/k-1}\rho_{\alpha^{2m}}(x_{ik+1}^{ik+k})&\le&
\hH_{\mbox{\tiny csw}}(X_{m+1}|X^m)
+\frac{1}{m}+\frac{2(m+1)\alpha^{m+1}}{n}\log\frac{n}{m}\nonumber\\
&\le&\frac{1}{m}\sum_{q=1}^{m}\hH_{\mbox{\tiny csw}}(X_q|X^{q-1})+\frac{1}{m}+
\frac{2(m+1)\alpha^{m+1}}{n}\log\frac{n}{m}\nonumber\\
&=&\frac{\hH_{\mbox{\tiny
csw}}(X^{m})}{m}+\frac{1}{m}+\frac{2(m+1)\alpha^{m+1}}{n}\log\frac{n}{m},
\end{eqnarray}
which yields
\begin{equation}
\frac{k}{n}\sum_{i=0}^{n/k-1}\rho_{\mbox{\tiny
LZ}}(x_{ik+1}^{ik+k})\le \frac{\hH_{\mbox{\tiny
csw}}(X^{m})}{m}+\frac{1}{m}+\frac{2(m+1)\alpha^{m+1}}{n}\log\frac{n}{m}+\epsilon_2(k,\alpha^{2m}),
\end{equation}
thus completing the proof of Theorem \ref{ub}.\\

\section{Chain Rule}
\label{chainrule}

Before presenting the chain-rule results, we first need to provide some
additional background, which is associated with conditional LZ compression.

\subsection{Background on Conditional LZ Compression}

In \cite{Ziv85}, the concept of LZ complexity was extended to account for finite-state lossless 
compression with side information, leading to the conditional version of LZ complexity. Given sequences 
$x^n$ and $y^n$, we apply the incremental parsing procedure of the LZ algorithm to the paired sequence 
$((x_1,y_1),(x_2,y_2),\ldots,(x_n,y_n))$.
As previously noted, this procedure ensures that all parsed phrases are distinct, 
except possibly for the final phrase, which may be incomplete. Let 
$c(x^n,y^n)$ denote the resulting number of distinct phrases.
For instance,\footnote{This example is taken from \cite{Ziv85}.} if
\begin{eqnarray}
x^6&=&0~|~1~|~0~1~|~0~1|\nonumber\\
y^6&=&0~|~1~|~0~0~|~0~1|
\end{eqnarray}
we have $c(x^6,y^6)=4$.
Let us denote by $c(x^n)$ the resulting number of different phrases
of $x^n$, and denote by $x(l)$ the $l$-th different $x$--phrase,
$l=1,2,\ldots,c(x^n)$. In the running example, $c(x^6)=3$. Next, let us denote
the number of times $x(l)$ appears in the parsing of $x^n$ by
$c_l(y^n|x^n)$. Then, obviously,
$\sum_{l=1}^{c(x^n)} c_l(y^n|x^n)=
c(x^n,y^n)$. In our example, $x(1)=0$, $x(2)=1$, $x(3)=01$,
$c_1(y^6|x^6)=c_2(y^6|x^6)=1$, and $c_3(y^6|x^6)=2$. Now, the conditional LZ
complexity of $y^n$ given $x^n$ is defined as
\begin{equation}
\rho_{LZ}(y^n|x^n)\dfn\frac{1}{n}\sum_{l=1}^{c(x^n)}c_l(y^n|x^n)\log
c_l(y^n|x^n).
\end{equation}
In \cite{Ziv85} it was shown that $\rho_{LZ}(x^n|y^n)$ is the main term of the
compression ratio achieved by the conditional version of the LZ algorithm
described therein (see
also \cite{UK03}), i.e., the length function, $LZ(x^n|y^n)$, of the coding scheme
proposed therein is upper bounded (in parallel to (\ref{lz-clogc})) by
\begin{equation}
\label{conditional-lz}
LZ(y^n|x^n)\le n\rho_{LZ}(y^n|x^n)+n\epsilon_3(n),
\end{equation}
where $\epsilon_3(n)$ is a certain sequence that tends to zero uniformly as $n\to\infty$.
On the other hand, analogously to \cite[Theorem 1]{ZL78}, it was shown in
\cite{me00}, that $\rho_{LZ}(y^n|x^n)$ is also the main term of a lower bound
to the compression ratio that can be achieved by any finite-state encoder with
side information at both ends, provided that the number of states is not
too large, similarly as described above for the unconditional version, i.e.,
\begin{equation}
\rho_s(y^n|x^n)\ge\rho_{\mbox{\tiny LZ}}(y^n|x^n)-\epsilon_4(n,s),
\end{equation}
where $\epsilon_4(n,s)$ tends to zero uniformly as $n\to\infty$ for fixed $s$, 
and $\rho_s(y^n|x^n)$ is the $s$-state compressibility of $y^n$ given the side
information $x^n$ (available to both encoder and decoder), which is defined in
the same manner as the unconditional $s$-state compressibility, but under an
encoder model where the output function $f$ and the next-state function $g$
are fed by both $x_i$ and $y_i$ (in addition to the current state $z_i$) at each
time instant $i$  -- see \cite{me00} for
details. 

The results of the previous section can be readily extended to the conditional
case. In particular, we will be interested in the relations to the conditional
entropy induced by statistics of non-overlapping blocks:
\begin{equation}
\frac{\hH_{\mbox{\tiny nob}}(Y^m|X^m)}{m}\le\frac{q}{n}\sum_{i=0}^{n/q-1}
\rho_{\mbox{\tiny
LZ}}(y_{iq+1}^{iq+q}|x_{iq+1}^{iq+q})+\epsilon_3(q)+\Delta_m(\alpha^q\beta^q,\beta),
\end{equation}
and 
\begin{equation}
\frac{\hH_{\mbox{\tiny nob}}(Y^p|X^p)}{p}\ge\frac{r}{n}\sum_{i=0}^{n/r-1}
\rho_{\mbox{\tiny
LZ}}(y_{ir+1}^{ir+r}|x_{ir+1}^{ir+r})-\frac{1}{p}-\epsilon_4(r,\alpha^p\beta^p),
\end{equation}
where $m$, $p$, $q$ and $r$ are positive integers, $r$ and $q$ being divisors
of $n$.

\subsection{Chain Rule Theorem}

Our main result in this section is the following.
\begin{theorem}
\label{chainrulethm}
For every three positive integers $k$, $q$ and $r$, every positive integer $n$ that is a
multiple of $k$, $q$ and $r$, and every
$(x^n,y^n)\in\calX^n\times\calY^n$, we have:
\begin{enumerate}
\item Upper bound:
\begin{eqnarray}
\label{chainruleupperbound}
\frac{k}{n}\sum_{i=0}^{n/k-1}\rho_{\mbox{\tiny
LZ}}(x_{ik+1}^{ik+k},y_{ik+1}^{ik+k})
&\le&\frac{q}{n}\sum_{i=0}^{n/q-1}\left[\rho_{\mbox{\tiny
LZ}}(x_{iq+1}^{iq+q})+\rho_{\mbox{\tiny
LZ}}(y_{iq+1}^{iq+q}|x_{iq+1}^{iq+q})\right]+\nonumber\\
& &\Delta_m(\alpha^q,\alpha)+\epsilon_3(q)+\Delta_m(\alpha^q\beta^q,\beta)+
\frac{1}{m}+\epsilon_2(k,\alpha^m\beta^m).
\end{eqnarray}
\item Lower bound:
\begin{eqnarray}
\label{chainrulelowerbound}
\frac{k}{n}\sum_{i=0}^{n/k-1}\rho_{\mbox{\tiny
LZ}}(x_{ik+1}^{ik+k},y_{ik+1}^{ik+k})
&\ge&\frac{r}{n}\sum_{i=0}^{n/r-1}\left[\rho_{\mbox{\tiny
LZ}}(x_{ir+1}^{ir+r})+\rho_{\mbox{\tiny
LZ}}(y_{ir+1}^{ir+r}|x_{ir+1}^{ir+r})\right]-\nonumber\\
& &\frac{2}{p}-\epsilon_2(r,\alpha^p)-\epsilon_4(r,\alpha^p\beta^p)-\Delta_p(\alpha^k\beta^k,\alpha\beta)-\epsilon_1(k).
\end{eqnarray}
\end{enumerate}
\end{theorem}

For the upper bound, the integer parameters $m$ and $q$ ($q$ being a divisor
of $n$) are free and can be chosen so
as to minimize the right-hand side. In particular, to make all redundancy
terms at the second line of eq.\ (\ref{chainruleupperbound}) small, the regime
should be $k\gg m\gg q\gg 1$, which means that we may upper bound the average
joint LZ complexity in terms of its decomposition, provided that the block
length $q$ of the blocks after the decomposition is very small relative to
the original block length, $k$. Likewise,
for the lower bound, the integer parameters $p$ and $r$ ($r$ being a divisor
of $n$) are free and can be chosen so
as to maximize the right-hand side. To make all redundancy
terms at the second line of eq.\ (\ref{chainrulelowerbound}) small, the regime
should be $r\gg p\gg k\gg 1$, which means that we can lower bound the average
joint LZ complexity in terms of its decomposition, provided that the block
length $r$ of the blocks after the decomposition is very large relative to
$k$. Combining both parts of Theorem \ref{chainrulethm}, the relevant regime is
therefore $r\gg p\gg k \gg m\gg q\gg 1$. In view of this, consider eq.\
(\ref{chainrulelowerbound}), take the limit superior of $n\to\infty$, then
the limit superior of $r\to\infty$, afterwards the limit of $p\to\infty$ and
finally, the limit of $k\to\infty$, to get
\begin{equation}
\rho(\bx,\by)\ge\limsup_{r\to\infty}\limsup_{n\to\infty}\frac{r}{n}\sum_{i=0}^{n/r-1}\left[\rho_{\mbox{\tiny
LZ}}(x_{ir+1}^{ir+r})+\rho_{\mbox{\tiny
LZ}}(y_{ir+1}^{ir+r}|x_{ir+1}^{ir+r})\right].
\end{equation}
On the other hand, consider eq.\ (\ref{chainruleupperbound}),
take the limit superior of $n\to\infty$, then
the limit of
$k\to\infty$, afterwards the limit of $m\to\infty$, and finally the limit
inferior of $q\to\infty$, to get
\begin{equation}
\rho(\bx,\by)\le\liminf_{q\to\infty}\limsup_{n\to\infty}\frac{q}{n}\sum_{i=0}^{n/q-1}\left[\rho_{\mbox{\tiny
LZ}}(x_{iq+1}^{iq+q})+\rho_{\mbox{\tiny
LZ}}(y_{iq+1}^{iq+q}|x_{iq+1}^{iq+q})\right].
\end{equation}
It follows that
\begin{equation}
\varrho_k(\bx,\by)=\limsup_{n\to\infty}\frac{k}{n}\sum_{i=0}^{n/k-1}\left[\rho_{\mbox{\tiny
LZ}}(x_{ik+1}^{ik+k})+\rho_{\mbox{\tiny
LZ}}(y_{ik+1}^{ik+k}|x_{ik+1}^{ik+k})\right],~~~~~k\in\calN
\end{equation}
is a convergent sequence whose limit $\varrho(\bx,\by)$ is equal to
$\rho(\bx,\by)$.

\noindent
{\em Proof of Theorem \ref{chainrulethm}.}
Let $(x^n,y^n)\in\calX^n\times\calY^n$ be a given pair of individual sequences. Then,
using the results of the previous section, we have the following relations.
For the upper bound,
\begin{eqnarray}
& &\frac{k}{n}\sum_{i=0}^{n/k-1}\rho_{\mbox{\tiny
LZ}}(x_{ik+1}^{ik+k},y_{ik+1}^{ik+k})\nonumber\\
&\le&\frac{\hH_{\mbox{\tiny
nob}}(X^m,Y^m)}{m}+\frac{1}{m}+\epsilon_2(k,\alpha^m\beta^m)\nonumber\\
&=&\frac{\hH_{\mbox{\tiny nob}}(X^m)}{m}+\frac{\hH_{\mbox{\tiny nob}}(Y^m|X^m)}{m}+\frac{1}{m}+
\epsilon_2(k,(\alpha\beta)^{2m})\nonumber\\
&\le&\frac{q}{n}\sum_{i=0}^{n/q-1}\rho_{\mbox{\tiny
LZ}}(x_{iq+1}^{iq+q})+\epsilon_1(q)+\Delta_m(\alpha^q,\alpha)+\nonumber\\
& &\frac{q}{n}\sum_{i=0}^{n/q-1}\rho_{\mbox{\tiny
LZ}}(y_{iq+1}^{iq+q}|x_{iq+1}^{iq+q})+\epsilon_3(q)+\Delta_m(\alpha^q\beta^q,\beta)+\nonumber\\
& &\frac{1}{m}+\epsilon_2(k,\alpha^m\beta^m)\nonumber\\
&=&\frac{q}{n}\sum_{i=0}^{n/q-1}\left[\rho_{\mbox{\tiny
LZ}}(x_{iq+1}^{iq+q})+\rho_{\mbox{\tiny
LZ}}(y_{iq+1}^{iq+q}|x_{iq+1}^{iq+q})\right]+\nonumber\\
& &\Delta_m(\alpha^q,\alpha)+\epsilon_3(q)+\Delta_m(\alpha^q\beta^q,\beta)+
\frac{1}{m}+\epsilon_2(k,\alpha^m\beta^m).
\end{eqnarray}
For the lower bound,
\begin{eqnarray}
& &\frac{k}{n}\sum_{i=0}^{n/k-1}\rho_{\mbox{\tiny
LZ}}(x_{ik+1}^{ik+k},y_{ik+1}^{ik+k})\nonumber\\
&\ge&\frac{\hH_{\mbox{\tiny
nob}}(X^p,Y^p)}{p}-\Delta_p(\alpha^k\beta^k,\alpha\beta)-\epsilon_1(k)\nonumber\\
&=&\frac{\hH_{\mbox{\tiny
nob}}(X^p)}{p}+\frac{\hH_{\mbox{\tiny nob}}(Y^p|X^p)}{p}-\Delta_p(\alpha^k\beta^k,\alpha\beta)-\epsilon_1(k)\nonumber\\
&\ge&\frac{r}{n}\sum_{i=0}^{n/r-1}\rho_{\mbox{\tiny
LZ}}(x_{ir+1}^{ir+r})-\frac{1}{p}-\epsilon_2(r,\alpha^p)+\nonumber\\
& &\frac{r}{n}\sum_{i=0}^{n/r-1}\rho_{\mbox{\tiny
LZ}}(y_{ir+1}^{ir+r}|x_{ir+1}^{ir+r})-\frac{1}{p}-\epsilon_4(r,\alpha^p\beta^p)-\nonumber\\
& &\Delta_p(\alpha^k\beta^k,\alpha\beta)-\epsilon_1(k)\nonumber\\
&=&\frac{r}{n}\sum_{i=0}^{n/r-1}\left[\rho_{\mbox{\tiny
LZ}}(x_{ir+1}^{ir+r})+\rho_{\mbox{\tiny
LZ}}(y_{ir+1}^{ir+r}|x_{ir+1}^{ir+r})\right]-\nonumber\\
& &\frac{2}{p}-\epsilon_2(r,\alpha^p)-\epsilon_4(r,\alpha^p\beta^p)-\Delta_p(\alpha^k\beta^k,\alpha\beta)-\epsilon_1(k).
\end{eqnarray}
This completes the proof of Theorem \ref{chainrulethm}.

\subsection{Comparison to an Earlier Derived Chain Rule}

In \cite{meSW}, the following chain-rule theorem was asserted and proved.
\begin{theorem}
Define
\begin{eqnarray}
\rho_{\mbox{\tiny LZ}}^+(x^k,y^k)&=&\max\{\rho_{\mbox{\tiny
LZ}}(x^k,y^k),\rho_{\mbox{\tiny LZ}}(x^k)+\rho_{\mbox{\tiny
LZ}}(y^k|x^k),\nonumber\\
& &\rho_{\mbox{\tiny LZ}}(y^k)+\rho_{\mbox{\tiny LZ}}(x^k|y^k)\},\\
\rho_{\mbox{\tiny LZ}}^-(x^k,y^k)&=&\min\{\rho_{\mbox{\tiny
LZ}}(x^k,y^k),\rho_{\mbox{\tiny LZ}}(x^k)+\rho_{\mbox{\tiny
LZ}}(y^k|x^k),\nonumber\\
& &\rho_{\mbox{\tiny LZ}}(y^k)+\rho_{\mbox{\tiny LZ}}(x^k|y^k)\}.
\end{eqnarray}
Given $\bx$ and $\by$, let
\begin{eqnarray}
\rho^+(\bx,\by)&=&\limsup_{k\to\infty}\limsup_{n\to\infty}\frac{k}{n}\sum_{i=0}^{n/k-1}\rho_{\mbox{\tiny
LZ}}^+(x_{ik+1}^{ik+k},y_{ik+1}^{ik+k})\nonumber\\
\rho^-(\bx,\by)&=&\limsup_{k\to\infty}\limsup_{n\to\infty}\frac{k}{n}\sum_{i=0}^{n/k-1}\rho_{\mbox{\tiny
LZ}}^-(x_{ik+1}^{ik+k},y_{ik+1}^{ik+k}).\nonumber
\end{eqnarray}
Then,
\begin{equation}
\label{equalrhos}
\rho^+(\bx,\by)=\rho^-(\bx,\by)=\rho(\bx,\by).
\end{equation}
\end{theorem}

This theorem tells that
upon dividing the infinite sequence into non-overlapping $k$-blocks,
then for the infinite sequence pair,
it does not matter if on each such block we apply LZ78 compression on $(x_{ik+1}^{ik+k},y_{ik+1}^{ik+k})$ jointly,
or first on $x_{ik+1}^{ik+k}$ and then on $y_{ik+1}^{ik+k}$ given $x_{ik+1}^{ik+k}$,
or vice versa, the ultimate compression ratio will be always the same.
However, in contrast to the chain-rule theorem presented here, which applies
for finite sequences (with clearly characterized redundancy terms), this theorem of
\cite{meSW} applies to infinite sequences only. Hence, the chain-rule theorem
presented here is more refined.

A natural question that may arise at this point is what can be said about the
relationship between $\rho(\bx,\by)$ and the pair ($\rho(\bx),\rho(\by|\bx)$)
(or ($\rho(\by),\rho(\bx|\by)$)). On the one hand, it is readily seen that
\begin{eqnarray}
& &\rho(\bx,\by)\nonumber\\
&=&\rho^-(\bx,\by)\nonumber\\
&\le&\limsup_{k\to\infty}\limsup_{n\to\infty}\frac{k}{n}\sum_{i=0}^{n/k-1}[\rho_{\mbox{\tiny
LZ}}(x_{ik+1}^{ik+k})+\rho_{\mbox{\tiny
LZ}}(y_{ik+1}^{ik+k}|x_{ik+1}^{ik+k})]\nonumber\\
&\le&\limsup_{k\to\infty}\limsup_{n\to\infty}\frac{k}{n}\sum_{i=0}^{n/k-1}\rho_{\mbox{\tiny
LZ}}(x_{ik+1}^{ik+k})+\nonumber\\
& &\limsup_{k\to\infty}\limsup_{n\to\infty}\frac{k}{n}\sum_{i=0}^{n/k-1}\rho_{\mbox{\tiny
LZ}}(y_{ik+1}^{ik+k}|x_{ik+1}^{ik+k})\nonumber\\
&=&\rho(\bx)+\rho(\by|\bx).
\end{eqnarray}
However, the reverse inequality,
$\rho(\bx,\by)\ge \rho(\bx)+\rho(\by|\bx)$,
may not\footnote{This corrects a
certain mistaken statement in \cite{meSW}.} hold true in general,
and so, there is no apparent chain rule in that sense.
As a counterexample, consider the following.
Let $n_0=0$ and $\{n_i,~i\ge 1\}$ be a sequence of positive integers
with the property that for all $i>1$, $n_i\gg\sum_{j=1}^{i-1}n_j$ and consider
an infinite binary sequence $\bx$, defined as follows: For $i$ even
and all $n_i+1\le t\le n_{i+1}$, $x_t=0$. For $i$ odd
and all $n_i+1\le t\le n_{i+1}$,
$x_t$ is obtained by random coin tossing.
Since $n_i\gg\sum_{j=1}^{i-1}n_j$, the
compression rate of the last segment always dominates, and so, the
compression rate of $x^n$
oscillates forever between $0$ and $1$, as $n$ grows without bound, which
results in a limit superior of $\rho(\bx)=1$ almost surely.
Next, let $\by$ be defined
as follows. For $i$ even and all $n_i+1\le t\le n_{i+1}$, $y_t$ is obtained by
independent random coin
tossing. For $i$ odd and all $n_i+1\le t\le n_{i+1}$, we set $y_t=x_t$.
Then, $\rho(\by|\bx)$ is the limit superior of a
sequence of conditional compression rates that oscillates between $0$ and $1$, and hence $\rho(\by|\bx)=1$,
implying that $\rho(\bx)+\rho(\by|\bx)=2$. On the other hand, when
compressing $(\bx,\by)$ jointly, in each segment the required
compression ratio is essentially
one bit per symbol pair, $(x_t,y_t)$, in all segments.
Therefore, $\rho(\bx,\by)=1$.
It is therefore apparent that the inequality
between $\rho(\bx,\by)$ and $\rho(\bx)+\rho(\by|\bx)$
stems mainly from the
limit superior operation and the possibility that $\rho(\bx)$ and
$\rho(\by|\bx)$ may be attained by different subsequences.

\end{document}